# Magneto-transport and Shubnikov-de Haas oscillations in the layered ternary telluride Ta$_3$SiTe$_6$ topological semimetal

Muhammad Naveed[1], Fucong Fei[1,*], Haijun Bu[1], Xiangyan Bo[1], Syed Adil Shah[1], Bo Chen[1], Yong Zhang[1], Qianqian Liu[1], Boyuan Wei[1], Shuai Zhang[1], Chuanying Xi[2], Xiangang Wan[1], Fengqi Song[1,†]

[1]*National Laboratory of Solid State Microstructures, Collaborative Innovation Center of Advanced Microstructures, and College of Physics, Nanjing University, Nanjing 210093, China*

[2]*High Magnetic Field Laboratory, Chinese Academy of Science, Hefei 230031, Anhui, China*

# ABSTRACT

Topological semimetals characterize a novel class of quantum materials hosting Dirac/Weyl fermions. The important features of topological fermions can be exhibited by quantum oscillations. Here we report the magnetoresistance and Shubnikov-de Haas (SdH) quantum oscillation of longitudinal resistance in the single crystal of topological semimetal $Ta_3SiTe_6$ with the magnetic field up to 38 T. Periodic amplitude of the oscillations reveals related information about the Fermi surface. The fast Fourier transformation spectra represent a single oscillatory frequency. The analysis of the oscillations shows the Fermi pocket with a cross-section area of 0.13 $Å^{-2}$. Combining magneto-transport measurements and the first-principles calculation, we find that these oscillations come from the hole pocket. Hall resistivity and the SdH oscillations recommend that $Ta_3SiTe_6$ is a hole dominated system.

# I. INTRODUCTION

Topological materials have been fascinating massive attention in the recent research of condensed matter physics [1-5]. The band structures in these materials reveal topology or symmetry shielded band crossing close to the Fermi energy so that the quasiparticles with low energy act very distinctly from the traditional types of fermions, which leads to unique physical characteristics. The most renown examples are Dirac and Weyl semimetals owning to four-fold and two-fold degenerate band crossing points accordingly. The electrons around these points resemble the relativistic Dirac and Weyl particles [6-18], giving birth to some exciting effects, for instance, extremely high mobility, large negative magnetoresistance and chiral anomaly [19, 20]. Apart from zero-dimensional (0D) modes in a three-dimensional (3D) system, there is a possibility that the nontrivial band crossing also takes the shape of one-dimensional (1D) nodal lines or two-dimensional (2D) nodal surfaces [3, 21-23]. These materials show quite amazing properties like anisotropic electron transport, drumheads like states and surface magnetism or superconductivity, etc [24-29].

There are two types of band crossing based on the formation procedure, accidental and essential band crossing. The first kind of band crossings is generated in particular regions of the Brillouin zone due to band inversions, which can be detached without disturbing the symmetry. The essential band crossing formed by a specific space group symmetry like in $Ta_3SiTe_6$ (nonsymmorphic space-group symmetry) which cannot be eliminated as far as the symmetry is sustained [30, 31]. Besides these essential band crossings in Dirac semimetals or nodal line materials, this was also found that nonsymmorphic symmetries can give surge to other exotic kinds of band crossing such as nodal chains and hourglass dispersions [32-36]. The latest theoretical investigation recommended that in the omission of spin orbit coupling (SOC) $Ta_3SiTe_6$ bulk crystals host accidental Dirac loop and four-fold nodal line, and an hourglass loop in the presence of SOC [37]. Regarding the research process on this material, the experimental studies are extremely rare and there is no report on the transport behavior of this material yet. Transport measurement is a useful technique to reveal the band structure and topological properties by analyzing the low-temperature and high-magnetic field transport phenomenon such as SdH oscillations, which have been described as a convincing tool for characterizing quantum transport in materials presenting 3D bulk and 2D surface states respectively [38,39].

Here we present the magneto-transport (MR) measurement and SdH quantum oscillation is observe in the single crystal of topological semimetal $Ta_3SiTe_6$. We perform low temperatures and high magnetic field magneto-transport analysis to produce experimental proof of quantum transport in $Ta_3SiTe_6$ crystal. We observe that at the low magnetic field the Hall resistivity support transport dominated carriers are hole-type. By increasing the magnetic field SdH quantum oscillations appeared above 28 T, the origin of these oscillations is also hole pocket because the cross-sectional area of the Fermi surface obtained from the experiments is almost consistent with the hole pocket we got from the theoretical calculations.

## II. EXPERIMENTAL SECTION

High-quality $Ta_3SiTe_6$ crystals were grown by chemical vapor transport (CVT) method using iodine as a transport agent. Highly pure powder of tantalum, silicon, and tellurium (from Alfa Aesar 99.98 %, 99.99%, 99.99%) were sealed in an evacuated quartz tube and put in a two-zone furnace. The furnace temperature was set at 950 °C to the source side and 850 °C to the growth side and the furnace was kept for seven days, and then naturally cooled down to room temperature. The energy-dispersive X-ray spectroscopy (EDS) and X-ray diffraction (XRD) were used to confirm the composition and structure of the single crystals.

Atomic Force Microscopy (AFM) measurements were performed by an Asylum Research Scanning Probe Microscope (SPM) operated in the AC mode, where an area of 30 μm by 30 μm around the flakes was scanned. Raman measurements were performed using a confocal Raman Microscope (Horiba XloRA) with 532 nm excitation laser, the power of the laser was kept low (200 μW) with the laser sport of

1 μm in diameter. The electrical transport measurements were carried out with a Physical Property Measurement System (DynaCool, Quantum Design) at Nanjing University, and High Magnetic Field Measurement Laboratory at Science Island Hefei China.

$Ta_3SiTe_6$ electronic structure calculations were carried out by using the full potential linearized augmented plane-wave method as implemented in WIEN2K package [40]. The General Gradient Approximate (GGA) for the correlation potential was used here. Using the second-order variational procedure [41], we included SOC interaction. The lattice constant we used was $a$ = 6.329 Å, $b$ = 11.414 Å and $c$ = 14.019 Å with space group Pnma [42]. The muffin-tin radii for Ta, Si and Te were set to 1.32, 1.10 and 1.32 Å. The basic functions were expanded to $R_{mt}R_{max}$ = 7 (where $R_{mt}$ is the smallest of the muffin-tin sphere radii and $R_{max}$ is the largest reciprocal lattice vector used in the plane-wave expansion). A 9 × 5 × 4 $k$-point mesh was used for the Brillouin zone integral. The self-consistent calculations are considered to be converged when the difference in the total energy of the crystal does not exceed 0.01 $mR_y$.

## III. RESULTS AND DISCUSSION

$Ta_3SiTe_6$ is a layered ternary telluride compound with space group No.62 (Pnma). The crystal structure of $Ta_3SiTe_6$ is composed of the stacks of sandwich layers which are similar to that of $MoS_2$ [43], as shown in Fig. 1(a). In $MoS_2$ each S-M-S sandwiched layer is formed of border sharing trigonal $MoS_6$ prisms [44]. But in the case of $Ta_3SiTe_6$ Te-(Ta-Si)-Te sandwich layers contain the face and edge-sharing $TaTe_6$ prism and the Si ions are placed into the interstitial sites between these prisms. The brown spheres represent Te, green Ta and blue Si respectively. EDS and XRD confirm the successful growth of the $Ta_3SiTe_6$ with perfect atomic ratio, layered structure and excellent crystallinity, as shown in Fig. 1(b, c). The crystals of the prepared sample are presented in the inset of Fig. 1(b), which clearly indicates the regular shiny surfaces of the crystals. The Raman spectrum is shown in Fig. 1(d) the most prominent peaks are at 78 cm$^{-1}$, 109.2 cm$^{-1}$, 133.2 cm$^{-1}$, 143 cm$^{-1}$, 148.8 cm$^{-1}$ and 156.7 cm$^{-1}$. As reported before that MoS$_2$ exhibited only E$_{2g}^1$ and A$_{1g}$ phonon modes [43], while Ta$_3$SiTe$_6$ displays more modes which probably associated with its relatively complicated lattice structure. We are unable to specify these extra phonon modes because of the lack of theoretical studies of phonon spectra. Layered material can be exfoliated as shown in the inset of Fig. 2(a), due to the micaceous nature of the Ta$_3$SiTe$_6$ it can be thinned down to one-unit cell thick 2D crystal by micro-exfoliation [45] and can offer quite different transport properties compared to the bulk. The micro-exfoliation techniques had made it possible to produce crystals with a different thickness that was formerly unreachable, and this has importantly increased the age of the research on low dimensional physics.

Fig. 2(a) represents temperature-dependent longitudinal resistivity with zero magnetic fields. The temperature dependence resistivity shows the metallic behavior under room temperature across the whole temperature range. The MR is measured at different temperatures with the magnetic field applied perpendicular, a semi-classical quadratic field-dependent behavior was observed, as shown in Fig. 2(b). The MR is defined as $[\rho(B)-\rho(0)]/\rho(0)\times 100\%$ where $\rho(B)$ and $\rho(0)$ is the resistivities at fields $B$ and zero respectively. The MR value reaches to 64% at 14 T and 2 K. As the temperature increased the MR is decreasing, the rate of scattering raised and the carrier mobility started getting reduced, quite small MR of 4.7% was observed at 14 T and 100 K. In order to examine the transport properties further, we measured the Hall resistivity at different temperatures. The resistivity as a function of the magnetic field is given in Fig. 2(c). The slope of the Hall coefficient remains positive during the entire measured temperature range at different temperatures which means that the hole carriers have dominated the transport [46, 47]. The value of the carrier density was $n = 5.2\times10^{21}$ cm$^{-3}$ at 2 K calculated by the formula $n = \dfrac{1}{R_H e}$.

For the sake of more information about the electronic structure, we have also investigated longitudinal resistivity under a high magnetic field. Initially, we observed the same behavior up to a certain magnetic field as recorded earlier at the low magnetic field, however above 28 T, we got the SdH oscillations described in Fig. 2(d). When the Landau levels pass through Fermi level of the system, which outcomes in the oscillation of the electronic density of states at the Fermi level. This anomaly creates oscillations in many electronic properties of the material such as resistance (SdH oscillation) and magnetization (de Haas-van Alphen effect) which are known as quantum oscillations. The frequencies of these oscillations are proportional to the cross-sectional area of the Fermi surface, perpendicular to the direction of the magnetic field [48]. When the higher magnetic field applied we observed SdH quantum oscillations as mentioned above. In order to study the oscillations in details, one can see from Fig. 3(a) which shows SdH oscillations at different temperatures, as the temperature raises the amplitudes of the oscillations decrease and oscillations were disappeared for 10K . Fig. 3(b) inset shows the corresponding oscillatory component $\Delta R_{xx}$ vs $\frac{1}{B}$. Furthermore Fig. 3(b) also shows the pronounced single oscillatory frequency F=1383T , which was determined from the fast Fourier transformation (FFT) analysis of the data, the value of the frequency is comparatively high so in such a situation it becomes very hard to tell the exact nature of the Berry phase (trivial, or non-trivial). The frequency directly gives the cross-sectional area of the Fermi surface by using Onsager relation $F = \left(\frac{\hbar}{2\pi e}\right) A_F$ where $A_F$ is the Fermi surface cross-sectional area, the value of the cross-sectional area was calculated as $A_F$ = 0.13 Å$^{-2}$ and the related wave factor was $k_F$ = 0.204 Å$^{-1}$. In order to evaluate the effective mass, we used Lifshitz-kosevich (LK) formula and fitted the temperature-dependent FFT amplitude [49].

$$\Delta M \propto -B^\lambda R_T R_D \sin\left\{2\pi - \left[\frac{F}{B} - \left(\frac{1}{2} - \phi\right)\right]\right\} \quad (1)$$

$R_T = \frac{\chi T}{\sinh(\chi T)}$ and $R_D = \exp(-\chi T_D)$ is the thermal damping factor and Dingle

damping factor respectively, the $\chi$ can be represented as $\chi = \dfrac{2\pi^2 k_B m^*}{\hbar e B}$, where $k_B$ indicates the Boltzmann constant and $m^*$ is the cyclotron mass. In equation (1) $\phi$ representing the phase shift, and $\phi$ can be denoted as $\phi = \dfrac{\varphi_B}{2\pi} - \delta$ where $\varphi_B$ is the Berry phase and $\delta$ have the values 0 and ±1/8 for 2D and 3D systems. When we took the oscillation amplitude of the peak B = 31 T, the cyclotron mass of the carriers $m^*$ was extracted to be 1.4 $m_e$ by conducting a theoretical fit with the above equation Fig. 3(c). The Fermi velocity was calculated by $v_F = \dfrac{\hbar k_F}{m^*}$ and got the value $v_F$ = 15.3 × 10⁵ m/s the Fermi energy $E_F$ = 58.2 meV was calculated by $E_F = m^* v_F^2$. The Dingle temperature in Fig. 3(d) is connected to the quantum-scattering lifetime $\tau$ which was obtained 14.5 K based on equation (1). The quantum-scattering lifetime $\tau$ = 8.3 × 10⁻¹⁴ s was calculated by $T_D = \dfrac{\hbar}{2\pi k_B \tau}$. We also obtained quantum mobility $\mu_q$ = 104.2 cm²V⁻¹s⁻¹ by using the formula $\mu_Q = \dfrac{e\tau}{m^*}$. The current theoretical study proposed that in the absence of SOC, the $Ta_3SiTe_6$ host the Dirac loop and mandatory fourfold nodal line and an hourglass Dirac loop in the presence of SOC. In order to search for the possible proof of the relativistic fermions in the materials, the study of the Berry phase is required. For Dirac materials, the pseudospin rotation under magnetic field should outcome in a non-trivial Berry phase, which can be achieved from the Landau-level (LL) index fan diagram or from the direct fit of the SdH oscillation pattern to the Lk formula [46]. In our case, it's quite hard to decide either Berry phase is trivial or non-trivial because the oscillatory frequency was comparatively high.

In order to find the origin of the SdH quantum oscillations, we carried out theoretical calculations. The band structure of $Ta_3SiTe_6$ including SOC is given in Fig. 4(a), where many characteristics can be observed. For example, each band is at least two-fold degenerate due to the presence of time-reversal and inversion symmetries.

The nodal loop around the Γ point is gapped because of SOC, each band is fourfold degenerate along the paths U-X, R-U, and Z-S, and more interesting is the emergence of hourglass-type dispersion on S-R. Such hourglass dispersion also appears on the S-X path [37]. Fig. 4(b, c) shown zoom-in images of the low-energy bands along S-R, one can see the hourglass dispersion around S-R high symmetry points. Fig. 4(d) describes the Brillouin zone of the 3D bulk, the high-symmetry points are marked. Fig. 4(e) illustrates the overall image of the Fermi pockets, among which two-hole like pockets were found, with cross-sectional area quite parallel to the area of the pocket we got from the experiments. These Fermi pockets are separately shown in Fig. 4(f, g). The cross-sectional area (marked by the red bold dashed lines) of these two pockets are between 0.084 $Å^{-2}$ and 0.104 $Å^{-2}$ which are close to the value 0.13 $Å^{-2}$ of the Fermi pocket obtained from the experiments, confirming the experimental and theoretical calculation consistency. It is expected that these SdH oscillations originated from the second hole pocket, because the cross-sectional area of the second pocket is comparatively more closely related to the experimental area of the pocket. For deeper insight into SdH oscillations, further experimental and theoretical studies will be needed.

## IV. CONCLUSION

In summary, we have performed magneto-transport measurement on a single crystal of the topological semimetal $Ta_3SiTe_6$ under magnetic field up to 38 T and low temperatures. The FFT spectra reveal the existence of the single oscillatory frequency. Hall resistivity demonstrates hole dominated transport at the low magnetic field, as the magnetic field increased SdH quantum oscillations were observed. Combining magneto-transport measurements and the first-principles calculation, we find that the origin of these oscillations is hole pocket because the experimental and theoretical calculated cross-sectional area of the pockets are consistent with each other.

## ACKNOWLEDGMENTS

The authors gratefully acknowledge the financial support of the National Key R&D Program of China (2017YFA0303203), the National Natural Science Foundation of China (91421109, 91622115, 11522432, 11574217, U1732273, U1732159, 11904165, and 11904166), the Natural Science Foundation of Jiangsu Province (BK20160659), the Fundamental Research Funds for the Central Universities, and the opening Project of Wuhan National High Magnetic Field center.

Corresponding authors : [*]feifucong@nju.edu.cn, [†]songfengqi@nju.edu.cn

# References

[1] C.-K. Chiu, J. C. Teo, A. P. Schnyder, and S. Ryu, Rev.Mod. Phys. 88, 035005 (2016).

[2] A. A. Burkov, Nat. Mater. 15, 1145-1148 (2016).

[3] S. A. Yang, SPIN 06, 1640003 (2016).

[4] X. Dai, Nat. Phys. 12, 727 (2016).

[5] A. Bansil, H. Lin, and T. Das, Rev. Mod. Phys. 88,021004 (2016).

[6] X. Wan, A. M. Turner, A. Vishwanath, and S. Y.Savrasov, Phys. Rev. B 83, 205101 (2011).

[7] S. Murakami, New J. Phys. 9, 356 (2007).

[8] A. A. Burkov and L. Balents, Phys. Rev. Lett. 107,127205 (2011).

[9] S. M. Young, S. Zaheer, J. C. Y. Teo, C. L. Kane, E. J.Mele, and A. M. Rappe, Phys. Rev. Lett. 108, 140405 (2012).

[10] Z. Wang, Y. Sun, X.-Q. Chen, C. Franchini, G. Xu, H. Weng, X. Dai, and Z. Fang, Phys.Rev. B 85, 195320 (2012).

[11] Z. Wang, H. Weng, Q. Wu, X. Dai, and Z. Fang, Phys.Rev. B 88, 125427 (2013).

[12] Y. X. Zhao and Z. D.Wang, Phys. Rev. Lett. 110, 240404 (2013).

[13] B.-J. Yang and N. Nagaosa, Nat. Commun. 5, 4898 (2014).

[14] H. Weng, C. Fang, Z. Fang, B. A. Bernevig, and X. Dai, Phys. Rev. X 5, 011029 (2015).

[15] Z. K. Liu, B. Zhou, Y. Zhang, Z. J. Wang, H. M. Weng, D. Prabhakaran, S.-K. Mo, Z. X.Shen, Z. Fang, X. Dai, Z. Hussain, and Y. L. Chen, Science 343, 864 (2014).

[16] S. Borisenko, Q. Gibson, D. Evtushinsky, V. Zabolotnyy, B. Buchner, and R. J. Cava, Phys. Rev. Lett. 113,027603 (2014).

[17] B. Q. Lv, H. M. Weng, B. B. Fu, X. P. Wang, H. Miao, J. Ma, P. Richard, X. C. Huang, L. X. Zhao, G. F. Chen, Z. Fang, X. Dai, T. Qian, and H. Ding, Phys. Rev. X 5,031013 (2015).

[18] S.-Y. Xu, I. Belopolski, N. Alidoust, M. Neupane, G. Bian, C. Zhang, R. Sankar, G. Chang, Z. Yuan, C.-C. Lee, S.-M. Huang, H. Zheng, J. Ma, D. S. Sanchez, B. Wang, A. Bansil, F. Chou, P. P. Shibayev, H. Lin, S. Jia, and M. Z. Hasan, Science 349, 613

(2015).

[19]  H. B. Nielsen and M. Ninomiya, Phys. Lett. B 130, 389 (1983).

[20]  D. T. Son and B. Z. Spivak, Phys. Rev. B 88, 104412 (2013).

[21]  C. Zhong, Y. Chen, Y. Xie, S. A. Yang, M. L. Cohen, and S. Zhang, Nanoscale 8,7232 (2016).

[22]  Q.-F. Liang, J. Zhou, R. Yu, Z. Wang, and H. Weng, Phys. Rev. B 93, 085427 (2016).

[23]  T. Bzdusek and M. Sigrist, Phys. Rev. B 96, 155105 (2017).

[24]  K. Mullen, B. Uchoa, and D. T. Glatzhofer, Phys. Rev. Lett. 115, 026403 (2015).

[25]  A. A. Burkov, M. D. Hook, and L. Balents, Phys. Rev. B 84, 235126 (2011).

[26]  R. Yu, Z. Fang, X. Dai, and H. Weng, Front. Phys. 12, 127202 (2017).

[27]  Y. Wang and R. M. Nandkishore, Phys. Rev. B 95, 060506 (2017).

[28]  T. T. Heikkila, N. B. Kopnin, and G. E. Volovik, JETP Lett. 94, 233 (2011).

[29]  J. Liu and L. Balents, Phys. Rev. B 95, 075426 (2017).

[30]  S. M. Young and C. L. Kane, Phys. Rev. Lett. 115, 126803 (2015).

[31]  Y. X. Zhao and A. P. Schnyder, Phys. Rev. B 94, 195109 (2016).

[32]  Z. Wang, A. Alexandradinata, R. J. Cava, and B. A. Bernevig, Nature 532, 189 (2016).

[33]  J. Ma, C. Yi, B. Lv, Z. Wang, S. Nie, L. Wang, L. Kong,Y. Huang, P. Richard, P. Zhang, et al., Sci. Adv. 3, e1602415 (2017).

[34]  T. Bzdusek, Q. S. Wu, A. Ruegg, M. Sigrist, and A. A. Soluyanov, Nature 538, 75 (2016).

[35]  S.-S. Wang, Y. Liu, Z.-M. Yu, X.-L. Sheng, and S. A. Yang, Nat. Commun. 8, 1844 (2017).

[36]  Qinghui Yan, Rongjuan Liu, Zhongbo Yan, Nature Physics 14,461-464 (2018)

[37]  S. Li, Y. Liu, S.-S. Wang, Z.-M. Yu, S. Guan, X.-L. Sheng, Y. Yao, and S. A. Yang, Phys. Rev. B 97, 045131 (2018).

[38]  Chang, L. L., Sakaki, H., Chang, C. A. & Wsaki, L. Shubnikov-de Hass Oscillations in a Semiconductor Superlattice. Phys. Rev. Lett. 38, 1489–1493 (1997).

[39]  Gao, H. et.al.Quantized Hall effect and Shubniko-de-Hass Oscillation in highly doped $Bi_2Se_3$: Evidence for Layered Transport of bulk carriers. Phys.Rev.Lett.108, 216803 (2012).

[40] Blaha, P., Schwarz, K., Madsen, G. K. H., Kvasnicka, D. & Luitz, J. WIEN2K, An Augmented Plane Wave+ Local Orbitals Program for Calculating Crystal Properties (Karlheinz Schwarz, Technische Universitat

[41] 2.D. D. Koelling and B. N. Harmon, J. Phys. C 10, 3107 (1977).

[42] M. Evain, L. Monconduit, A. Van der Lee, R. Brec, J. Rouxel, and E. Canadell, New J. Chem. 18, 215 (1994)

[43] Lee, C.et.al. Anomalous Lattice Vibrations of Single and Few-Layer $MoS_2$. ACS Nano 4, 2695-2700 (2010).

[44] O. Lopez-Sanchez, D. Lembke, M. Kayci, A. Radenovic, and A. Kis, Nature nanotechnology 8, 497 (2013).

[45] J. Hu, X.Liu, Nature Physics 11, 471-476 (2015).

[46] Linlin An, Hongwei Zang, Physical Review B 97,235133 (2018)

[47] Wenhong Wang, Yin Du, Guizhou Xu, Scientific Reports 3, 2181 (201

[48] Arnab Kumar Pariari, arXiv preprint arXiv: 1905.06255 (2019)

[49] I. M. Lifshitz and A. M. Kosevich, Sov. Phys. JETP 2, 636 (1956).

# Figure Captions

**Fig. 1. Ta$_3$SiTe$_6$ crystal growth and characterization.** (a) The MoS$_2$ type crystal structure of Ta$_3$SiTe$_6$, brown sphere represents Te, green Ta, and blue Si respectively. (b) The EDS spectrum of the synthesized Ta$_3$SiTe$_6$ crystal, which demonstrates a stoichiometric ratio, inset is the optical image of the crystal. (c) The single-crystal X-ray diffraction of the $(00n)$ surfaces of the sample. (d) Raman measurements of the bulk.

**Fig. 2. Electrical transport under high magnetic fields and low temperatures.** (a) The resistivity varies with temperature at zero fields, inset is the AFM image of the bi-layer flake with different thickness, scanned area 30μm by 30μm. (b) The MR measured at different temperatures. (c) The positive linear Hall resistivity with an applied magnetic field at different temperatures. (d) Longitudinal resistivity with SdH oscillations as the magnetic field range increased, device representation in the inset.

**Fig. 3. SdH oscillations measurement and analysis.** (a) Amplitudes of the oscillations decrease as temperature rises. (b) The single frequency and extracted oscillatory component of the oscillation are given in the inset. (c) Taking the oscillation amplitude of the peak $B = 31 \text{T}$, the effective mass of the carrier extracted by performing a theoretical fit. (d) The Dingle temperature obtained by Lk formula.

**Fig. 4. Electronic band structure calculations of Ta$_3$SiTe$_6$.** (a) Energy band structure. (b, c) The zoom-in images of the low-energy band along S-R. (d) Brillouin zone of the 3D bulk. (e) Fermi surfaces overall view. (f, g) Calculated hole like pockets with different cross-sectional areas.

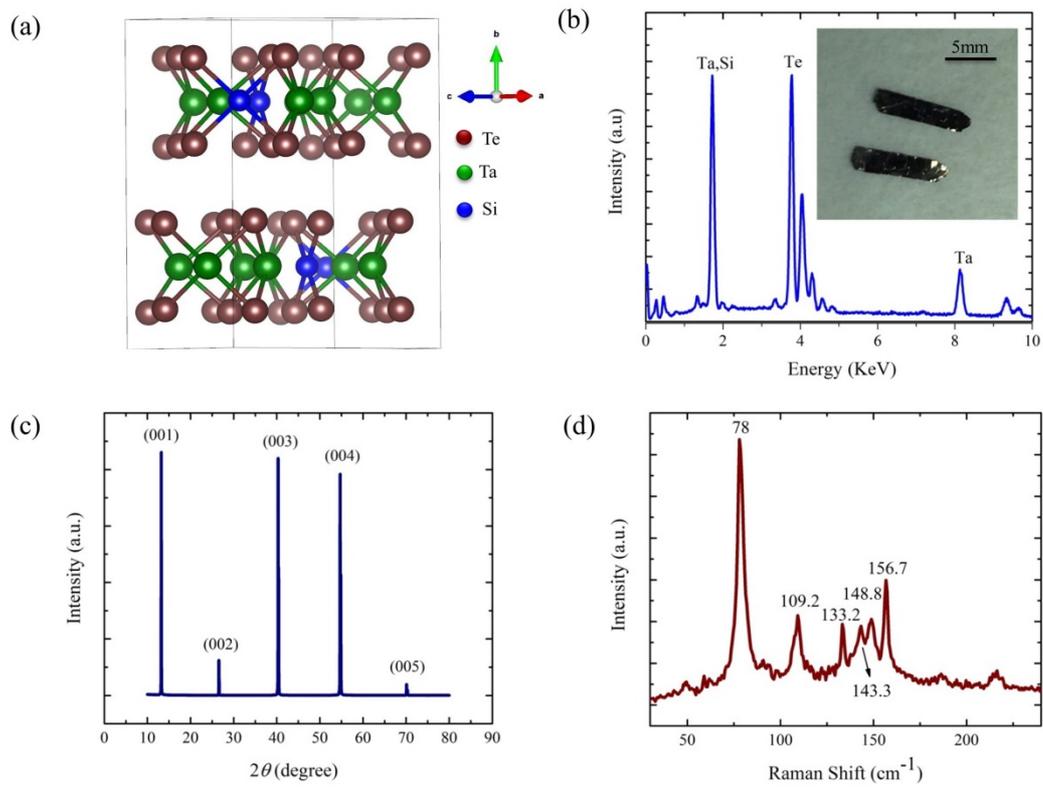

**Fig. 1.** Ta$_3$SiTe$_6$ **crystal growth and characterization.**

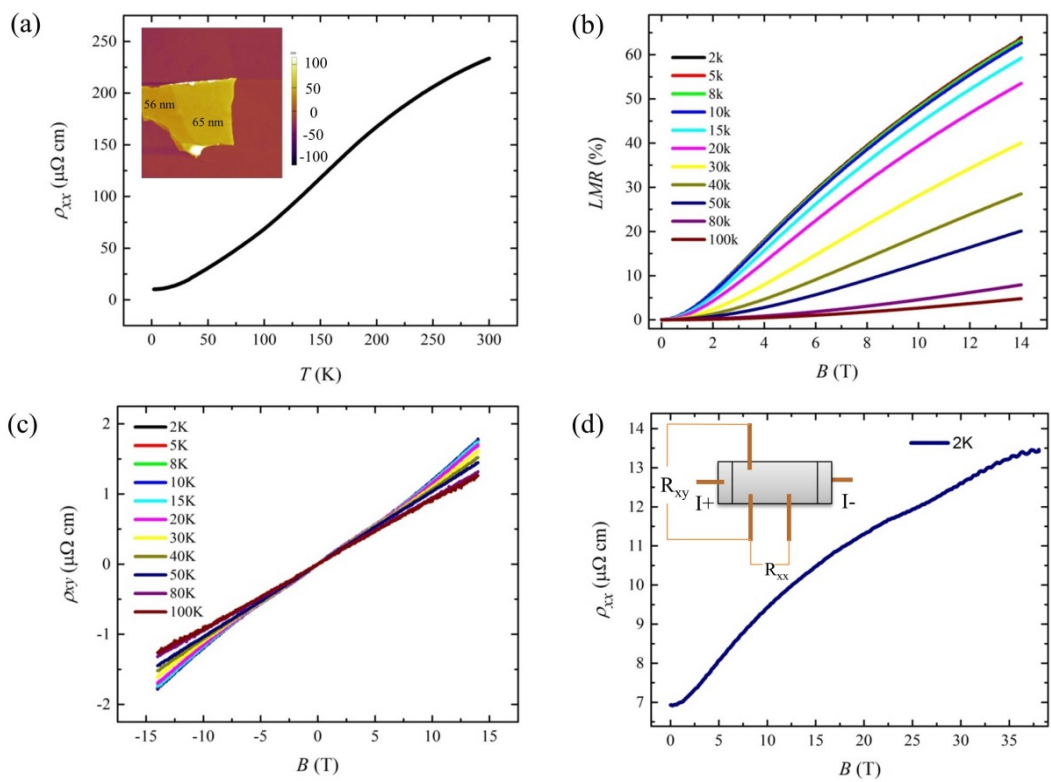

**Fig. 2. Electrical transport under high magnetic fields and low temperatures.**

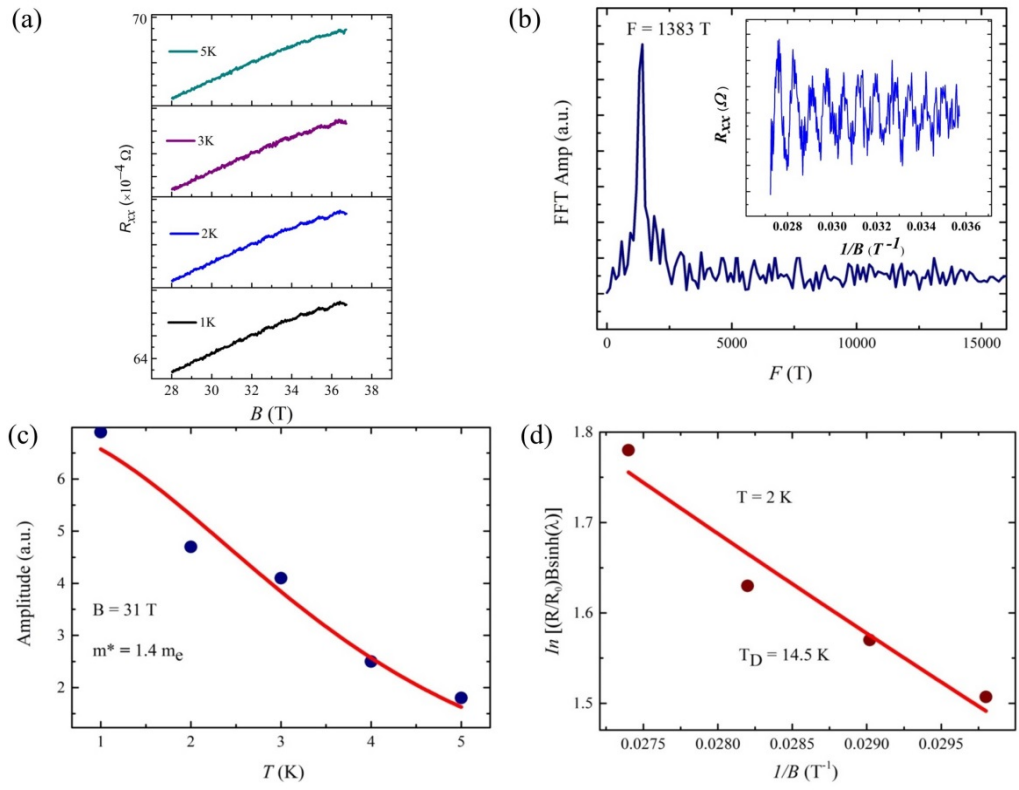

**Fig. 3. SdH oscillations measurement and analysis.**

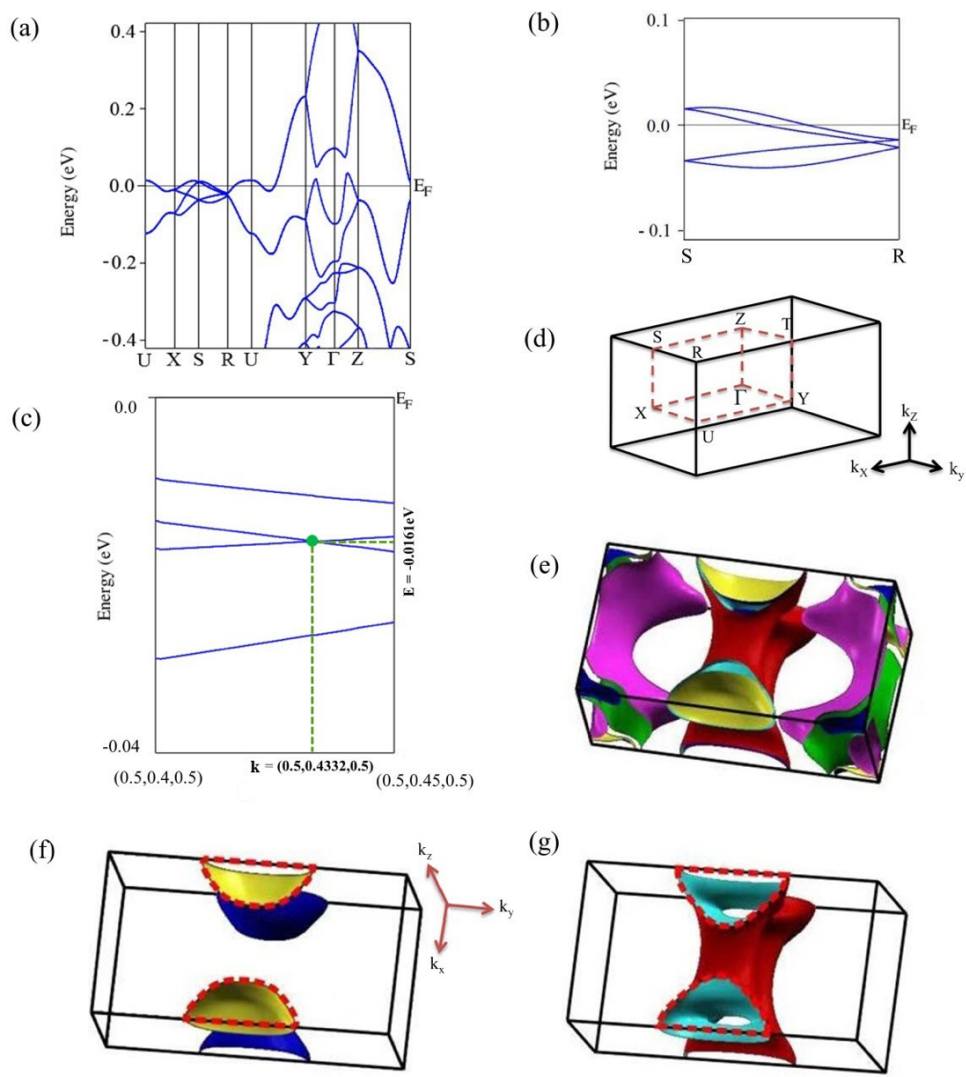

**Fig. 4. Electronic band structure calculations of $Ta_3SiTe_6$.**